\documentclass[lettersize,journal]{IEEEtran}
\usepackage[utf8]{inputenc}
\usepackage{textgreek}
\usepackage{amsmath,amsfonts,amssymb}
\usepackage{array,booktabs,multirow,tabularx}
\usepackage[caption=false,font=footnotesize,labelfont=sf,textfont=sf]{subfig}
\usepackage{textcomp}
\usepackage{stfloats}
\usepackage{url}
\usepackage{graphicx}
\usepackage{cite}
\usepackage{xcolor}
\usepackage{tikz}
\usepackage{enumitem}
\usetikzlibrary{arrows.meta, positioning, fit, shapes.geometric, calc}
\newcommand{\sysone}{System~1}
\newcommand{\systwo}{System~2}
\newcommand{\taskA}{Task~A}
\newcommand{\taskB}{Task~B}
\newcommand{\taskC}{Task~C}
\newcommand{\NTT}{\textsc{NTT}}
\newcommand{\ITT}{\textsc{ITT}}
\newcommand{\BC}{\textsc{BC}}
\newcommand{\BI}{\textsc{BI}}
\newcommand{\NA}{\textsc{NA}}
\newcommand{\projectrepo}{\protect\url{https://github.com/Ruddy202/DualTurn}}

\begin{document}
\bstctlcite{IEEEexample:BSTcontrol}

\title{Leveraging Dual-Process Theory for Multimodal Multiparty Turn-Taking Prediction}


\author{Rutherford A. Patamia, Ming Liu, Wei Luo, Favour Ekong, Rehan Mohammed, Gang Liu and Akan Cosgun%
\thanks{R. A. Patamia, M. Liu, W. Luo, and R. Mohammed are with the School of Information Technology, Deakin University, Melbourne, VIC, Australia. F. Ekong is with Griffith University, Brisbane, QLD, Australia. G. Liu is with the College of Computer Science and Technology, Harbin Engineering University, China. A. Cosgun is with Arm, Cambridge, United Kingdom. Corresponding authors: Rutherford A. Patamia
(e-mail: r.patamia@deakin.edu.au) and Ming Liu
(e-mail: m.liu@deakin.edu.au).

Project repository: \projectrepo}}


\maketitle

\begin{abstract}
In multiparty conversations, a pause does not always mean a speaker has finished, and another voice does not necessarily indicate a change in who holds the conversational floor. Reliable turn-taking therefore requires recognising not only whether a speaker is ready to yield the floor, but also when a possible transition deserves attention. We propose a dual-process-inspired framework that separates continuous candidate discovery from selective verification. A fast, recall-oriented stage identifies plausible transition points. In contrast, a deliberative stage combines acoustic, linguistic, speaker-related, and predicted future-activity evidence, together with multimodal cues where available, to reject false candidates and refine transition timing. We evaluate the framework through three distinct tasks: SHIFT/HOLD classification at supplied decision points, verification of automatically generated candidates, and continuous discovery and localisation of floor transfers. On VoxConverse, we derive action-aware turn-taking events from speaker-diarization annotations to examine unconstrained conversations with varying numbers of speakers. Multi-TPC provides complementary evaluation of controlled triadic interactions using speech, gaze, motion, and prosodic information. The findings reveal that identifying plausible transitions is considerably easier than reliably confirming and localising genuine floor transfers, while the contribution of individual evidence sources varies across conversational settings. These results highlight the importance of treating multiparty turn-taking as a continuous discovery-and-verification problem rather than classification at predefined boundaries.
\end{abstract}


\section{Introduction}
\IEEEPARstart{T}{urn-taking} coordinates who speaks, who listens, and when the conversational floor changes \cite{sacks1974}. Here, the \emph{floor} denotes the temporary opportunity to make the main conversational contribution. Multiparty interaction complicates this process: several listeners may compete for the floor, a listener may backchannel without taking over, an interruption may begin before the current speaker stops, and a pause may occur even though the same speaker intends to continue. Turn-taking is therefore a temporal and action-interpretation problem rather than simple silence detection.

Human turn gaps are often only a few hundred milliseconds, while spoken-response planning can take longer \cite{heldner2010,levinson2015}. Turn coordination also combines lexical completion, prosody, gaze, gesture, facial behaviour, and the activity of other participants \cite{kendrick2023multimodal,skantze2021}. A system that waits for a long silence reacts slowly, whereas one that responds after a short silence risks interrupting the speaker. Predictive models instead estimate likely near-future interaction from the available history.

Prior work established several parts of this problem. Continuous recurrent models predict future speech activity during pauses and overlap \cite{skantze2017,roddy2018continuous}; Voice Activity Projection (VAP) predicts joint near-future activity and supports continuous real-time inference \cite{vap,realtimevap}; TurnGPT adds linguistic completion evidence \cite{turngpt}; and Triadic VAP and MuVAP extend future-activity reasoning to multiparty settings \cite{triadicvap,muvap}. Controlled multimodal studies further show that interaction state, gaze, motion, and gesture can inform turn behaviour \cite{lee2023turn,heo2025gaze,lee2025motion,suresh2026gestures}, while language context can aid next-speaker reasoning \cite{fukuda2026llm}. The unresolved issue addressed here is not whether turn-taking can be predicted continuously, but how performance changes when the system must also discover \emph{where} a meaningful floor-transfer decision should be made.

This distinction matters because seemingly similar scores can arise under different information contracts. At a supplied utterance or inter-pausal-unit boundary, a model is told where to decide and only interprets the boundary. In continuous audio, it must find plausible moments, reject false opportunities, localise the surviving event, and distinguish genuine transfers from backchannels, brief interjections, overlap, and same-speaker continuation. We therefore separate three evaluation levels: \taskA{} is reference decision-point SHIFT/HOLD classification; \taskB{} verifies automatically generated candidates; and \taskC{} performs continuous end-to-end floor-transfer discovery. Task~C is the headline operational task, while Tasks~A and B diagnose where performance is gained or lost.

To address this separation, we apply dual-process theory as an \emph{event-triggered allocation of computation} \cite{dualprocess}. \sysone{} continuously scans for potential floor changes, prioritising coverage over precision, while only its candidates activate \systwo{}, which evaluates acoustic, linguistic, future-activity, speaker-related, and local competition evidence before finalising and timing an event. This approach highlights the distinct roles of candidate generation, selective verification, and localisation, each measured separately. The psychological analogy supports this division of labour but does not imply cognitive equivalence or end-to-end real-time operation.

The empirical study uses complementary datasets rather than treating all multiparty corpora as interchangeable. VoxConverse \cite{voxconverse} is an unconstrained, variable-speaker diarization corpus; we derive an action-aware turn-taking contract from its speaker-labelled regions while preserving the official development/test separation. Multi-TPC \cite{lee2026multitpc} provides controlled triadic interaction with synchronised speech, gaze, motion, and prosody. We preserve the original frozen Multi-TPC held-out result and separately report a post-hoc seven-group nested analysis so that single-split evidence is not presented as cross-group confirmation.

The main contributions of this work are:
\begin{itemize}[leftmargin=1.2em,itemsep=1pt,topsep=2pt]

    \item An explicit three-level evaluation contract separating supplied-point classification, automatic-candidate verification, and continuous event discovery, making missed opportunities, false proposals, and localisation errors distinguishable;

    \item An action-aware adaptation of VoxConverse that derives distinct conversational events with action-specific timestamps, canonicalised references, and human audit of derived labels;

    \item A selective dual-process-inspired architecture coupling recall-oriented candidate discovery with acoustic, linguistic, future-activity, and speaker-aware verification, ranking, and temporal localisation; and

    \item Complementary empirical evaluation across VoxConverse and multimodal Multi-TPC, using same-task baselines, group-disjoint held-out testing, post-hoc seven-group analysis, and modality ablations to examine robustness and limitations.

\end{itemize}

\section{Related Work}
\subsection{Continuous turn prediction and future activity}
Conversation analysis treats turn exchange as coordinated projection rather than a reaction to silence alone \cite{sacks1974}. Computational models likewise moved from endpoint rules toward continuous prediction: recurrent systems estimate future speech activity throughout interaction and around pauses and overlap \cite{skantze2017,roddy2018continuous}. VAP reframes the problem as self-supervised prediction of near-future joint voice activity \cite{vap} and has been demonstrated in continuous real-time operation \cite{realtimevap}. TurnGPT contributes complementary linguistic completion evidence \cite{turngpt}. These studies establish that turn-taking need not wait for a supplied endpoint. Our focus is the additional operational burden of generating sparse transition opportunities, verifying them, and localising a floor-transfer event in continuous multiparty recordings. Predicted future activity therefore serves as evidence for the decision rather than the event definition itself.

\subsection{Multiparty and multimodal turn-taking}
Multiparty interaction introduces several potential speakers and listener actions. Lee \emph{et al.} model three-party turn keeping and switching at predefined inter-pausal units using speech/backchannel state and gaze \cite{lee2023turn}; later work addresses online multimodal end-of-turn prediction and learned motion cues \cite{lee2024eot,lee2025motion}. Triadic VAP extends future-activity projection to three speakers \cite{triadicvap}, while MuVAP uses role-relative projection and multimodal speaker grounding to scale beyond fixed channel identities \cite{muvap}. Their task contracts differ from unconstrained event discovery, so published scores are used as context rather than as a cross-paper leaderboard; our Multi-TPC adaptations are instead evaluated on common outer groups and a common event contract.

Visual behaviour can resolve ambiguities that audio alone cannot. Gaze has been used for next-speaker and start-time prediction \cite{ishii2013gaze,ishii2016gaze}, and recent work examines multi-user gaze, spatial speaker information, visual VAP, and gesture-aware turn modelling \cite{heo2025gaze,russell2025visual,saga2025encoders,suresh2026gestures}. The reported benefits are event and setting-dependent, motivating our Multi-TPC ablations rather than assuming that every available modality must help.

\subsection{Speaker reasoning, agents, and positioning}
Predicting \emph{whether} the floor changes is distinct from predicting \emph{who} speaks next. Prior work uses gaze, linguistic context, and multimodal meeting information for next-speaker prediction \cite{ishii2016gaze,lee2024nsp,fukuda2026llm}. Our VoxConverse speaker diagnostic instead identifies the incoming speaker from post-boundary query speech; even the prefix-restricted inventory condition is therefore not causal next-speaker forecasting. Streaming conversational-agent work, such as ModeratorLM, further asks when an artificial participant should intervene \cite{mitra2026moderator}. That policy problem is downstream from the measurement problem considered here.

Dual-process theory has also informed dialogue planning, combining fast policies with deliberative search \cite{he2024dualplanning}. However, that work addresses conversational action selection rather than continuous multiparty turn-taking prediction. Our contribution is distinct in both its application and evaluation setting. Direct quantitative comparisons use common datasets and evaluation protocols; published cross-paper results provide context only.

\section{Problem Formulation}
\subsection{Floor-Transfer Event Formulation}
Let a conversation be represented by a continuous waveform $x(t)$ involving a set of speakers $\mathcal{S}$. A \emph{decision point} is a time at which the system must determine whether the current conversational floor is retained or transferred. We define \emph{SHIFT} as a genuine transfer of the main floor to a different speaker, whereas \emph{HOLD} denotes the absence of such a transfer. This distinction is not equivalent to speaker-change detection: brief listener responses, backchannels, and local overlap may introduce another voice without transferring the main floor.

At a supplied decision time $d_i$, the reference classification problem is
\begin{equation}
\hat{y}_i =
f\!\left(x_{\leq d_i}, z_{\leq d_i}\right),
\qquad
\hat{y}_i \in \{\mathrm{HOLD},\mathrm{SHIFT}\},
\end{equation}
where $z_{\leq d_i}$ denotes auxiliary evidence available no
later than $d_i$, including transcript history, speaker-activity
estimates, and cached past-context representations.
Accordingly, the decision is conditioned only on information
available by the specified decision time.

Continuous floor-transfer discovery removes the supplied decision point. The system must instead scan the interaction and produce an ordered set of predicted transition times,
\begin{equation}
\widehat{\mathcal{E}}
=
\{\hat{t}_1,\ldots,\hat{t}_M\}.
\end{equation}
Predicted and reference events are matched one-to-one within a temporal tolerance $\delta$. We use $\delta=0.50$~s as the primary criterion and report $0.70$ and $1.00$~s as localisation-sensitivity analyses. We compute precision, recall, and F1 from the resulting matched and unmatched events.

\subsection{Evaluation Tasks and Operational Difficulty}
\begin{figure*}[t]
    \centering
    \includegraphics[width=0.92\textwidth]{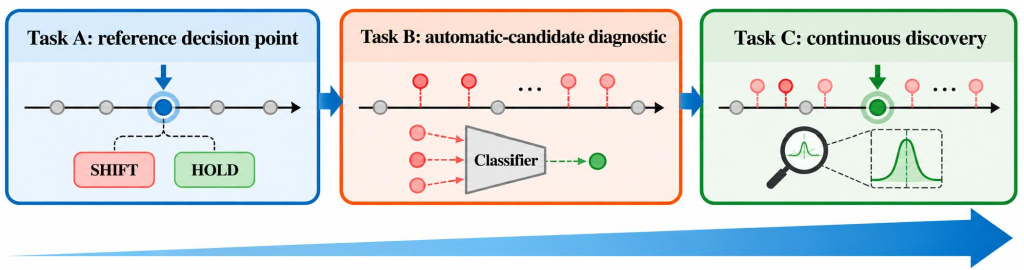}
    \caption{Evaluation decomposition for continuous multiparty turn-taking. 
    Task A evaluates SHIFT/HOLD classification at a supplied reference decision point. 
    Task B evaluates verification at automatically generated candidate points. 
    Task C evaluates continuous end-to-end transition discovery, including candidate proposal, verification, localisation, and event matching.}
    \label{fig:taskdecomp}
\end{figure*}

Figure~\ref{fig:taskdecomp} illustrates the three evaluation levels, each progressively diminishing the information available to the evaluator. 

\taskA{} is the cleanest diagnostic. For the main setup, the decision is made 100~ms after the end of the previous speaker segment and uses only the permitted history. We also evaluate alternative points 500~ms and 1000~ms before the endpoint to measure how much performance depends on proximity to the boundary.

\taskB{} evaluates the same underlying decision problem at automatically generated candidate times. The classifier is trained on the feature subset shared between reference decision points and automatic candidates, then applied to the noisier proposal distribution produced by the candidate-generation stage. This task therefore measures the system's ability to reject plausible but incorrect transition opportunities. It remains a diagnostic task and is distinct from the complete Task~C inference route.

\taskC{} is an end-to-end continuous detection task. The process begins with \sysone{} generating a high-recall candidate pool, followed by \systwo{} verifying and ranking the candidates. The system produces final event sets through validation-locked temporal post-processing and local timing correction. These sets are then evaluated against canonical reference events via one-to-one temporal matching. Task~C therefore requires the system to determine both \emph{where} to consider a floor-transfer decision and \emph{whether} the candidate represents a genuine transfer.

\subsection{Next-Speaker Identification}
For matched floor-transfer events, we evaluate incoming-speaker
identification using a short post-boundary audio segment.
Even when speaker prototypes are restricted to past enrollment
audio, the query contains speech from the incoming speaker.
Consequently, this experiment measures the effects of speaker-inventory
and enrollment assumptions rather than causal next-speaker
prediction.

\section{Repurposing VoxConverse for Turn-Taking}
\subsection{Dataset Characteristics and Suitability}
VoxConverse was introduced as a large-scale audio-visual speaker diarization dataset collected from unconstrained web video \cite{voxconverse}. Its original task is to determine which person speaks at each time, not to label floor-transfer actions. The development set contains 216 videos and approximately 1,218 minutes, while the test set contains 232 videos and approximately 2,612 minutes. The original paper reports a mean of 4.5 speakers per development recording and 6.5 per test recording, with maxima of 20 and 21 speakers, respectively. The recordings include debates, panel discussions, interviews, talk shows, and other in-the-wild material, and the original annotations include overlapping speech. A separate post-hoc language analysis estimated that English accounted for 98.98\% of speaker-attributed speech in DEV and 56.55\% in TEST, revealing a marked difference in language composition between the partitions.

These properties are attractive for multiparty turn-taking because the recordings are diverse and the number of speakers is not fixed. They also create a methodological problem. A diarization segment boundary does not automatically mean that a conversational turn has changed. VoxConverse annotation guidelines split speech around pauses, and the same speaker can therefore appear in adjacent segments separated by a gap. A different speaker can also produce a short backchannel or brief interjection that should not be counted as acquisition of the main floor. We therefore treat the diarization annotation as raw temporal evidence from which a separate turn-taking ontology must be derived.

The official TEST inventory is never used for model or threshold selection. The 216 development recordings are divided once into 177 TRAIN and 39 VAL sessions. All model selection, threshold selection, cooldown settings, and local correction parameters are fixed using TRAIN/VAL before TEST evaluation.

\subsection{Action-Aware Event Ontology}

For each temporally adjacent pair of diarised speech regions, we derive a turn-taking action from speaker identity, gap or overlap, and incoming-segment duration. Table~\ref{tab:ontology} summarises the resulting labels and their temporal anchors.

\begin{table*}[t]
\caption{Action-aware conversion of adjacent diarization segments into turn-taking events. SHIFT denotes sustained main-floor transfer, whereas short listener responses remain HOLD.}
\label{tab:ontology}
\centering
\scriptsize
\begin{tabularx}{\textwidth}{p{1.3cm}p{2.4cm}p{2.0cm}p{1.4cm}p{2.3cm}X}
\toprule
Action & Plain-language interpretation & Structural condition & Class & Primary event time & Reason for separate treatment \\
\midrule
\NA & No acquisition / same-speaker continuation & Same speaker appears again after a boundary or pause & HOLD & Previous speaker end & A pause or segmentation boundary is not itself a floor transfer. \\
\NTT & Normal turn transfer & Different incoming speaker, no qualifying overlap, incoming segment $>0.50$~s & SHIFT & Previous speaker end & The current speaker releases the floor before the next speaker's main contribution. \\
\ITT & Interruption turn transfer & Different incoming speaker, overlap $\geq0.06$~s, incoming segment $>0.50$~s & SHIFT & Incoming speaker onset & The new speaker begins before the old speaker stops, so onset is the meaningful acquisition point. \\
\BC & Backchannel & Different speaker, overlap $\geq0.06$~s, incoming segment $\leq0.50$~s & HOLD & Incoming speaker onset & Brief listener feedback can overlap without taking the main floor. \\
\BI & Brief interjection & Different speaker, no qualifying overlap, incoming segment $\leq0.50$~s & HOLD & Incoming speaker onset & A short contribution should not automatically be counted as a sustained floor transfer. \\
\bottomrule
\end{tabularx}
\end{table*}

The incoming-segment duration threshold is 0.50~s, the minimum qualifying overlap is 0.06~s, and positive-gap normal transfers are restricted to an immediate-transition window of at most 1.0~s. These values define the task contract and are not selected to optimise TEST performance.

The temporal anchor depends on the action type. \NTT{} is referenced to the release of the previous speaker's floor, whereas \ITT{} is referenced to the incoming speaker's onset because the takeover begins before the previous speaker has stopped.

\subsection{Canonicalization and label-quality safeguards}
The raw TEST conversion contains 6,350 SHIFT rows, including two exact duplicates at the same session and timestamp. Canonicalization removes duplicate \texttt{(session, time)} keys, yielding 6,348 unique TEST SHIFT events used consistently in all secondary analyses. Because the event ontology is derived algorithmically from diarization rather than native turn-taking annotations, we manually audited 300 short audio clips, balanced as 60 examples each of \NTT, \ITT, \BC, \BI, and \NA. The derived fine-grained action agreed with the human judgment in 226/300 cases (75.3\%; 95\% Wilson CI: 70.2--79.9\%; Cohen's $\kappa=0.697$). Agreement was strongest for \BI{} and \NTT{}, while most disagreements involved the more ambiguous distinction among interruptions, backchannels, and brief interjections. Collapsing the ontology to floor transfer (\NTT/\ITT) versus non-transfer (\BC/\BI/\NA) increased agreement to 86.0\%. Human confidence was high (mean 4.59/5; 89.7\% of judgments rated 4 or 5). No alternative event times were recorded, so the audit supports the action-label and floor-transfer distinctions rather than a separate quantitative validation of event timing.

\section{Dual-Process Framework}
\subsection{Functional interpretation}

\begin{figure*}[t]
    \centering
    \includegraphics[width=0.95\textwidth]{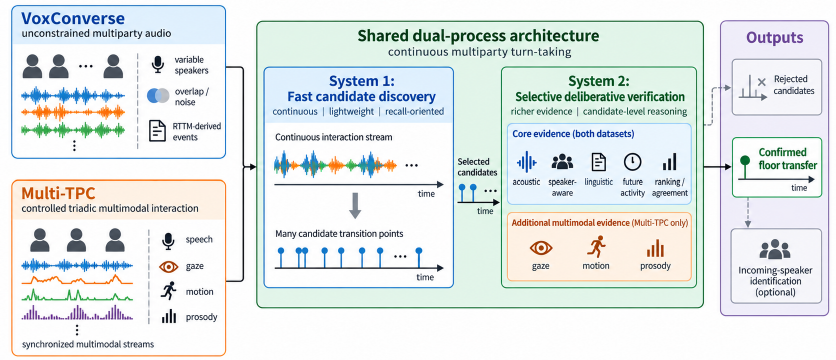}
    \caption{Shared dual-process-inspired architecture for continuous multiparty turn-taking. 
    VoxConverse provides an unconstrained, audio-centric evaluation with variable speaker counts and diarization-derived events, whereas Multi-TPC provides controlled triadic interaction with synchronised speech, gaze, motion, and prosodic information. 
    System~1 continuously performs fast, recall-oriented candidate discovery. 
    System~2 is selectively invoked at candidate times and combines richer acoustic, speaker-aware, linguistic, future-activity, ranking, and, where available, multimodal evidence to reject false candidates and confirm floor transfers. 
    Incoming-speaker identification is evaluated as a downstream post-boundary task.}
    \label{fig:framework}
\end{figure*}

Figure~\ref{fig:framework} illustrates how the two processes divide an otherwise continuous decision problem. \sysone{} asks \emph{when a floor change might warrant examination}; it scans densely and passes a high-recall set of candidate times forward. \systwo{} asks \emph{whether a particular candidate is a genuine transfer and where it occurred}; it is invoked selectively, using evidence unavailable to a simple silence trigger. A dense deliberative module operating at 10~Hz would be invoked 600 times per minute. At candidate rate \( r_c \), the approximate selective cost is

\begin{equation}
C_{\mathrm{dual}} \approx C_1 + r_c C_2,
\end{equation}
where \( C_1 \) represents the ongoing proposal cost and \( C_2 \) denotes the deliberative cost per candidate. The design is advantageous only if \sysone{} achieves high event coverage while maintaining \( r_c \) significantly below the dense frame rate. Accordingly, we present candidate ceiling recall and candidates per minute alongside the final F1 score.

\subsection{System 1: Recall-oriented Candidate Discovery}  
The primary acoustic proposer employs WavLM \cite{wavlm} as its pretrained speech representation, utilising a dense multi-scale temporal convolutional head. Audio is resampled to 16 kHz, and the proposer uses 4.0 s candidate-centred audio windows at a 0.10 s hop. During training, the head learns an end-of-turn-like score according to the action-aware event contract. At inference, local peaks are extracted and merged. The proposer prioritises retaining genuine transitions for recovery by \systwo{}, rather than acting as a high-precision event classifier. Therefore, the validation policy prioritises candidate recall above a specified threshold, with precision and F1 scores serving as secondary tie-breaking criteria. Additional candidates are derived from automatic diarization and ASR-generated boundaries. Candidate-source agreement is considered a feature, as support from multiple weak proposal mechanisms can lend greater credibility to a time than a peak derived from a single source.

\subsection{Temporal Verification}
The temporal verifier examines acoustic context around
automatically proposed candidates to distinguish genuine floor
transfers from pauses, hesitation, noise, and repeated peaks.
The executed VoxConverse pipeline includes bounded
post-candidate information, so its evaluation represents
offline verification rather than strictly past-only streaming
inference. Because temporal cues alone cannot reliably
distinguish speaker continuation from floor acquisition,
System~2 combines this evidence with linguistic,
future-activity, and speaker-change cues. Subsequent
candidate ranking and local timing correction help reject
false proposals and refine transition localisation.

\subsection{Future-Activity Evidence}
Future Activity Projection (FAP) predicts whether the current speaker will continue speaking and whether another participant may become active, building on the motivations outlined in VAP \cite{vap}. During training, the model utilises past context to anticipate future activity bins for both the current and other speakers. Notably, future activity serves solely as a training target and is not provided during TEST inference. For bins \( b=1,\ldots,B \), the model estimates:
\begin{equation}
\hat{p}^{\mathrm{cur}}_b = P(A^{\mathrm{cur}}_b=1\mid H_t), \qquad
\hat{p}^{\mathrm{oth}}_b = P(A^{\mathrm{oth}}_b=1\mid H_t),
\end{equation}
where $H_t$ represents the information available up to time $t$.
A decrease in expected activity for the current speaker,
coupled with an increase for another speaker, suggests a
potential floor transfer, while sustained activity from the
current speaker indicates HOLD.

In addition to this general framework, \systwo{} employs role-relative FAP (RR-FAP) to assess candidate-level evidence for the current floor holder rather than fixed speaker channels. RR-FAP integrates acoustic, linguistic, and diarization-derived features to determine which candidate within a local transition area most aligns with a floor transfer. Reference event information is used exclusively for constructing TRAIN/VAL targets, and TEST inference does not access future activity labels or ground-truth event data. Consequently, FAP and RR-FAP metrics measure predictive activity and candidate discrimination, distinct from the one-to-one event-detection F1 scores reported for Task~C.

\subsection{Linguistic Completion and Local Language-Model Evidence}

Automatic transcription is conducted using a Whisper-family
recogniser \cite{whisper}. A lightweight Qwen3 model
\cite{qwen3}, enhanced through low-rank fine-tuning, extracts
linguistic evidence surrounding each candidate. This approach
helps distinguish sentence-completion-like boundaries from
hesitations or syntactically incomplete continuations, while
also capturing action-relevant context not easily represented
by short acoustic windows. In the VoxConverse dataset, ASR
errors, fillers, fragmentary speech, laughter, and overlapping
speech are prevalent, making linguistic completion insufficient
to distinguish genuine floor transfers from phenomena such as
backchanneling. Consequently, the final model incorporates
the language branch as one of several specialised sources
of evidence.

\subsection{Speaker Evidence, Candidate Ranking, and Timing Correction}
Speaker-change evidence is obtained from automatic diarization
cues and Emphasized Channel Attention, Propagation and
Aggregation in Time Delay Neural Network (ECAPA-TDNN)
embeddings \cite{ecapa}. Diarization captures local speaker
continuity or change, while ECAPA similarity assesses speaker
characteristics around a candidate. Although informative,
these signals alone are insufficient; the silence-plus-speaker-change
baseline improves precision but has limited recall, especially
during interruptions without mutual silence.
System~2 integrates diverse acoustic scores, probabilities, source indicators, gap/overlap estimates, speaker cues, and language-derived features using validation-tuned tree-based models, such as LightGBM \cite{lightgbm}. A local event ranker treats nearby candidates as competing explanations for the same transition, reducing duplicate detections associated with a single event rather than applying independent thresholds to each peak and relying mainly on cooldown suppression.

The final operating threshold for Task~C, along with local gap constraints, ranking configuration, and timing-correction parameters, is established using validation data prior to the TEST evaluation. For accepted candidates, a validation-trained local correction predicts a minor temporal offset. This correction does not expand the evaluation tolerance or alter ground-truth event times; it corrects systematic candidate misalignment under the original event-matching criterion.

\subsection{Incoming-Speaker Identification}

For matched floor-transfer events, we use ECAPA-style speaker
prototypes to identify the incoming speaker under three conditions:
reference-based RTTM speaker inventories, automatic full-recording
diarization, and automatic speaker prototypes constructed from
speech preceding the decision time. These conditions differ in
their reliance on reference annotations and available speaker
history. All three use a post-boundary query segment from the
incoming speaker and therefore evaluate speaker identification
after a transition rather than causal next-speaker prediction.

\section{Experimental Protocol}
\subsection{Complementary Multi-TPC Dataset}

Multi-TPC complements the in-the-wild VoxConverse evaluation with controlled triadic interactions containing synchronised speech, gaze, motion, and prosody \cite{lee2026multitpc}. The published corpus comprises 24 sessions from seven three-participant groups (approximately 5.3~h); the locally audited release used here contains 23 complete sessions from the same seven groups. Unlike VoxConverse, whose turn-taking events are derived from RTTM diarization, Multi-TPC provides participant-level activity on a native 60-Hz timeline with inactive, primary-speech, and backchannel states. We align audio using verified offset-only correction with the time scale fixed at 1.0, without affine temporal warping. Session \texttt{mtpc\_20220211\_s02} has a severe motion-duration mismatch, so we exclude its motion stream rather than artificially realigning it.
The datasets serve complementary roles: VoxConverse evaluates continuous turn-taking under variable speaker counts and unconstrained acoustic conditions, whereas Multi-TPC supports controlled multimodal and cross-group analysis. Section~\ref{sec:multitpc} presents the Multi-TPC event contract, evaluation protocols, comparators, and results. Experimental notebooks, configuration files, and reproducibility documentation are available in the project repository.

\subsection{Reproducibility Configuration}

Table~\ref{tab:config} summarises the frozen VoxConverse
settings that define the experiment, including the acoustic
proposer, temporal verifier, ASR/language branch, speaker
evidence, and event-matching contract \cite{wavlm,qwen3,ecapa}.
Model and operating-point selection uses only TRAIN/VAL data,
with TEST reserved for final evaluation. Additional implementation
details and reproducibility documentation are available in the
project repository.

\begin{table*}[t] 
\caption{Frozen implementation settings that define the VoxConverse experiment.} 
\label{tab:config} 
\centering 
\scriptsize 
\begin{tabularx}{\textwidth}{p{2.4cm}p{3.1cm}X} 
\toprule 
Component & Frozen Setting & Experimental Role \\ 
\midrule 
Randomness / Audio & Seed 101; 16~kHz mono & Fixed context windows; TEST data excluded from selection. \\ 
System~1 & WavLM Large (Base+ fallback); 4.0~s context; 0.10~s hop & Dense scan followed by local-peak proposals. \\ 
System~1 Training & 35 epochs; batch size of 8; LR $2\times10^{-4}$ & Recall-oriented VAL selection. \\ 
Temporal Verifier & 0.10~s hop; 45 epochs; LR $8\times10^{-4}$ & Effective batch size capped at 128. \\ 
ASR / Language & Faster-Whisper medium; Qwen3-1.7B & Max length 512; LoRA rank 16/alpha 32; 3 epochs; batch size of 4. \\ 
Speaker Evidence & Pyannote + ECAPA & Automatic change cues and prototype similarity. \\ 
Event Matching & $\pm0.50$~s primary & Localization sensitivity analyses at $\pm0.70$ and $\pm1.00$~s. \\ 
\bottomrule 
\end{tabularx} 
\end{table*} 

\subsection{Same-Data Continuous Baselines}

We assess two continuous baselines on the VoxConverse TEST set, using canonical ground truth and a $\pm0.50$~s event-matching criterion for a controlled comparison. The \emph{automatic silence only} baseline identifies mutual-silence intervals based on automatic speaker activity, predicting SHIFT after a minimum silence determined during validation. No ground-truth timestamps are utilised for predictions; VAL specifies a minimum silence of 0.10~s.
The \emph{silence plus automatic speaker change} baseline requires a positive internal mutual-silence interval followed by a different automatically diarised speaker. VAL sets a maximum silence of 1.50~s and a 1.00~s cooldown. While this approach enables the rejection of same-speaker pauses via speaker change, it remains structurally limited in detecting interruptions that begin during the current speaker's active phase, which do not contain a mutual-silence interval.

\subsection{Augmentation campaign}
The controlled augmentation study evaluates four conditions: no augmentation, synthetic multiparty augmentation only, background-bed mixing only, and their full combination. During training, each waveform undergoes one or two randomly selected transformations, with one transformation chosen with a probability of 0.6. The transformations include gain scaling (0.85--1.15), Gaussian noise (18--30~dB signal-to-noise ratio), temporal shifts (up to $\pm10$~ms), speed perturbation (0.98--1.02), and mild low or high-pass filtering.
Background-bed mixing is applied independently with a probability of 0.6, using one of 96 pre-generated 16-kHz audio beds. The mixing signal-to-noise ratio is sampled from 10--25~dB. A deterministic seed selects the bed segment; segments shorter than the target waveform are repeated as necessary. The selected segment is Root Mean Square (RMS)-scaled with a signal floor of 0.02, mixed with the training waveform, and clipped to $[-1,1]$.

\subsection{Evaluation Metrics and Uncertainty}

For the reference decision-point classification, we report several metrics: accuracy, SHIFT precision/recall/F1, macro-F1, balanced accuracy, AUROC, and AUPRC. For continuous event discovery, we perform one-to-one temporal matching to obtain true positives (TP), false positives (FP), and false negatives (FN), from which we compute precision, recall, and F1 scores. The candidate ceiling recall represents the fraction of ground-truth SHIFT events that have at least one System~1 proposal within the primary tolerance before System~2 rejection.
We estimate uncertainty through session-level bootstrap resampling, using 1,000 replicates for the primary Task~C confidence interval and 2,000 for paired augmentation comparisons. Aggregate precision, recall, and F1 are recalculated from resampled TP, FP, and FN counts. The main Task~C F1 is reported with a 95\% confidence interval. For same-task system comparisons, paired session bootstrap is employed to ensure corresponding recordings are resampled together. A paired performance difference whose 95\% confidence interval excludes zero suggests that the observed difference is not solely attributable to session-level sampling variability. 
Runtime measurements are confined to the benchmarked System 1 core and the cached System 2 temporal verifier; all uncached processes, including diarization, ASR, Qwen3 inference, ECAPA extraction, and feature construction, are excluded.

\section{VoxConverse Results}
We present a concise overview of our results, starting with the performance on primary tasks. We compare decision-point classification, automatic candidate verification, and continuous discovery, followed by an evaluation of same-data baselines, System~1 candidate coverage, and localisation tolerance. Additionally, we analyse factors such as component interactions, event types, augmentation strategies, speaker inventories, interaction complexity, and runtime to clarify the observed end-to-end performance.

\subsection{Performance Across Evaluation Tasks}
Table~\ref{tab:mainresults} reveals a significant performance gap across the evaluated tasks. Task~A demonstrates the highest performance when supplied with the decision point, whereas Task~B shows the lowest results with automatically proposed candidates verified independently. Task~C improves upon Task~B through candidate verification, ranking, and localization, yet remains considerably below Task~A. This performance differentiation is attributed to the nature of the tasks: Task~A focuses on SHIFT/HOLD interpretation at a known decision time, while Task~C requires identifying plausible transition locations in continuous audio prior to confirming and localising a floor transfer. The observed difference between Task~A SHIFT F1 and Task~C event F1 is 0.313 (95\% CI [0.294, 0.331]), indicating a clear separation in task difficulty rather than a decline in model performance.

\begin{table*}[t]
\caption{Primary VoxConverse TEST results for Tasks A--C.
Task~A reports Macro-F1 and SHIFT-class precision and recall;
Tasks~B and C report event-level metrics.}
\label{tab:mainresults}
\centering
\scriptsize
\setlength{\tabcolsep}{4.0pt}
\renewcommand{\arraystretch}{1.15}
\begin{tabularx}{\textwidth}{@{}cXlllrrrrX@{}}
\toprule
Task &
Setting &
Decision point &
Primary metric &
Score &
95\% CI &
P &
R &
Supporting evidence \\
\midrule

A &
Supplied-point classification &
Supplied &
Macro-F1 &
0.849 &
-- &
0.874 &
0.858 &
SHIFT F1 = 0.866; TP/FP/FN = 4400/636/731 \\

B &
Candidate verification &
Proposed &
Event F1 &
0.455 &
[0.434, 0.478] &
0.441 &
0.471 &
TP/FP/FN = 2988/3790/3360 \\

C &
Continuous discovery &
Discovered &
Event F1 &
0.552 &
[0.530, 0.573] &
0.536 &
0.569 &
TP/FP/FN = 3609/3122/2739 \\

\bottomrule
\end{tabularx}
\end{table*}

\subsection{Continuous Same-Data Baselines}

\begin{figure}[t]
\centering
\includegraphics[width=\columnwidth]{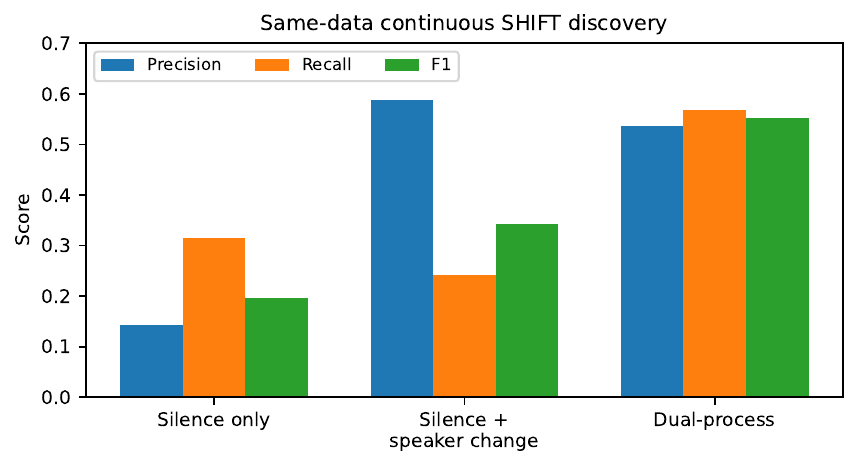}
\caption{Precision, recall, and F1 for the two automatic continuous baselines
and the final dual-process system.}
\label{fig:baselines}
\end{figure}

The results in Table~\ref{tab:baselines} and Fig.~\ref{fig:baselines} compare the systems on the same canonical TEST set under the $\pm0.50$~s continuous event-matching criterion. Automatic silence-based leads to a high number of false positives due to within-speaker pauses and segmentation gaps. Requiring an automatic speaker change substantially improves precision but decreases recall. In contrast, the dual-process system achieves a better precision-recall balance. Paired session bootstrap supports these improvements, with 95\% confidence intervals for F1 differences notably excluding zero.
Silence-based baselines are structurally limited for interruptions, as they open decision opportunities only after mutual silence. This limitation results in an ITT recall of 5.9\% for automatic silence and approximately 3.2\% when speaker change is also required. In contrast, the proposed system can generate candidates during or around overlap, offering a distinct advantage in recall performance.

\begin{table*}[t]
\caption{VoxConverse TEST continuous floor-transfer discovery at ±0.50 s. Paired ΔF1 denotes the F1 difference relative to the baseline identified in the final column.}
\label{tab:baselines}
\centering
\scriptsize
\setlength{\tabcolsep}{3.8pt}
\renewcommand{\arraystretch}{1.12}
\begin{tabularx}{\textwidth}{@{}XrrrrrrllX@{}}
\toprule
System &
P &
R &
F1 &
TP &
FP &
FN &
Paired $\Delta$F1 &
95\% CI &
Reference \\
\midrule
Automatic silence only &
0.142 &
0.314 &
0.196 &
1995 &
12026 &
4353 &
-- &
-- &
-- \\

Silence + automatic speaker change &
0.588 &
0.241 &
0.342 &
1529 &
1072 &
4819 &
$+0.146$ &
[0.117, 0.177] &
Automatic silence \\

Final dual-process &
0.536 &
0.569 &
\textbf{0.552} &
3609 &
3122 &
2739 &
$+0.210$ &
[0.187, 0.232] &
Silence + speaker change \\
\bottomrule
\end{tabularx}
\end{table*}

\subsection{Candidate Ceiling, Verification, and Localisation}
\begin{figure}[t]
\centering
\includegraphics[width=\columnwidth]{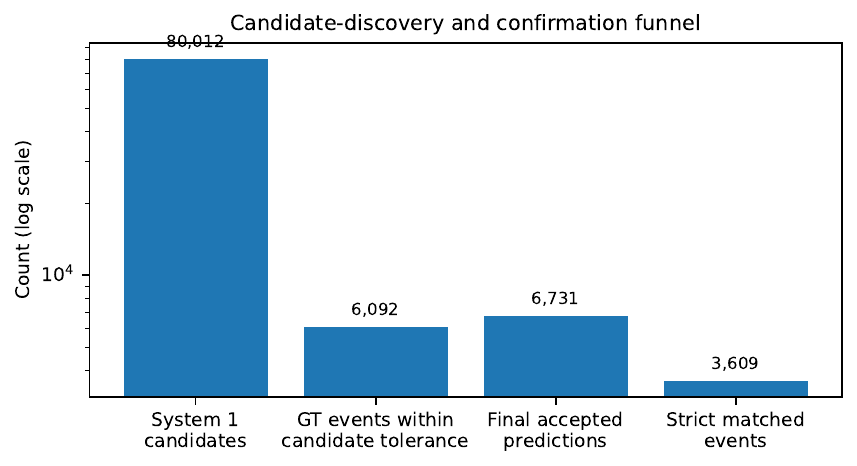}
\caption{Candidate-discovery and confirmation funnel on
VoxConverse TEST. System~1 achieves high reference-event
coverage, while System~2 rejects false candidates before
final event matching.}
\label{fig:funnel}
\end{figure}

Figure~\ref{fig:funnel} illustrates the candidate-discovery and confirmation process. System~1 identifies a large number of TEST candidates, capturing a high proportion of SHIFT events with a ceiling recall of 0.960. However, System~2 eliminates about 91.6\% of these candidates, ultimately yielding 6,731 predictions and 3,609 strict matches. This indicates a critical bottleneck in accurately selecting and localising transitions, rather than merely generating candidates. Despite the substantial proposal coverage, final recall is significantly reduced, highlighting challenges in distinguishing correct transitions from plausible alternatives. The current ceiling of 0.960 limits potential improvements through increased proposal recall alone and suggests that denser candidate generation could lead to more false alternatives needing rejection. The focus remains on maintaining the correct local event while filtering out nearby incorrect candidates, a challenge central to the deliberative System~2 stage.

With the final prediction set fixed at 6,731 events, increasing the matching tolerance from $\pm0.50$ to $\pm0.70$ and $\pm1.00$~s raises Task~C F1, indicating that some remaining errors are temporal near misses rather than complete detection failures. The $\pm0.50$~s criterion remains the primary evaluation setting; the wider tolerances are reported only as localisation sensitivity analyses.

The sparse candidate set reduces System~2 invocations.
At 30.63 candidates/min, temporal-verifier calls decrease
by 94.9\% relative to dense 10-Hz evaluation.
The System~1 core achieves a real-time factor (RTF) of
0.232, while the cached temporal verifier takes
0.0113~ms per candidate. These component-level measurements
exclude diarization, ASR, Qwen3 inference, ECAPA extraction,
and feature construction, and therefore do not establish
end-to-end runtime.

\subsection{Component Route Comparison}
Table~\ref{tab:ablation} presents a comparison of VAL-tuned System~2 configurations evaluated under the same TEST event-matching criteria. The results indicate that the temporal action verifier alone demonstrates limited effectiveness, suggesting that local temporal evidence is inadequate for reliable floor-transfer confirmation. The binary candidate verifier, local event ranker, and RR-FAP selector yield comparable nominal F1 scores, highlighting that diverse candidate representations facilitate effective verification through varying decision frameworks. The full dual-process configuration achieves the highest nominal F1, reflecting a superior precision-recall balance rather than consistent metric improvements. Notably, the meta-confirmer fails to enhance performance compared to simpler verification methods, thereby not contributing independently to performance gains.

\begin{table}[t]
\caption{Independently VAL-tuned System~2 configurations on VoxConverse TEST.
Each standalone configuration uses its own VAL-selected operating point; the
full row is the locked final system.}
\label{tab:ablation}
\centering
\scriptsize
\setlength{\tabcolsep}{3.2pt}
\renewcommand{\arraystretch}{1.12}
\begin{tabularx}{\columnwidth}{@{}X X rrr@{}}
\toprule
Configuration & Decision function & P & R & F1 \\
\midrule
Temporal action verifier &
Action classification; SHIFT from NTT/ITT &
0.255 & 0.446 & 0.324 \\

Binary candidate verifier &
Candidate-level SHIFT/HOLD &
0.569 & 0.525 & 0.546 \\

Local event ranker &
Best local candidate for a SHIFT event &
0.546 & 0.542 & 0.544 \\

RR-FAP selector &
Role-relative best SHIFT candidate &
0.557 & 0.533 & 0.545 \\

Meta-confirmer &
Learned SHIFT/HOLD confirmation from fused evidence &
0.578 & 0.500 & 0.536 \\

Full dual-process &
VAL-locked temporal--ranker fusion with timing correction &
0.536 & 0.569 & \textbf{0.552} \\
\bottomrule
\end{tabularx}
\end{table}

\subsection{Future-Activity Diagnostics}

Table~\ref{tab:fap} reports future-activity diagnostics at the main
\texttt{silent\_100ms} reference points. FAP shows strong discrimination of
near-future speaker activity, with other-speaker activity predicted more
reliably than current-speaker activity. These diagnostics should not be
interpreted as floor-transfer detection: predicted activity from another
participant may correspond to a backchannel, brief interjection, or overlap
that does not result in floor acquisition. System~2 therefore uses future
activity as supporting evidence rather than equating predicted other-speaker
activity with SHIFT. A role-relative FAP variant is used within System~2 for
candidate-level discrimination; its event-level contribution is reported in
Table~\ref{tab:ablation}.

\begin{table}[t]
\caption{Future-activity diagnostics. FAP reports TEST
micro-averaged future-activity prediction at
\texttt{silent\_100ms} decision points, while RR-FAP reports
VAL candidate discrimination at a threshold of 0.50.
These metrics are not directly comparable with Task~C event F1.}
\label{tab:fap}
\centering
\scriptsize
\setlength{\tabcolsep}{3.2pt}
\renewcommand{\arraystretch}{1.12}
\begin{tabularx}{\columnwidth}{@{}Xlrrrr@{}}
\toprule
Diagnostic target & Split & Acc. & F1 & AUROC & AUPRC/AP \\
\midrule
FAP: all activity bins (micro) &
TEST & 0.834 & 0.809 & 0.905 & 0.869 \\

FAP: current-speaker bins (micro) &
TEST & 0.833 & 0.748 & 0.887 & 0.826 \\

FAP: other-speaker bins (micro) &
TEST & 0.835 & 0.846 & 0.908 & 0.892 \\

RR-FAP: role-relative candidate discrimination &
VAL & -- & 0.521 & 0.887 & 0.509 \\
\bottomrule
\end{tabularx}
\end{table}

\subsection{Event-Type Analysis}
At reference decision points, SHIFT recall is high for both NTT (0.865) and ITT (0.839), whereas HOLD rejection is substantially lower for BC (0.283) and BI (0.344) than for other/unspecified HOLD events (0.887). This indicates that brief listener speech near a decision point can resemble genuine floor acquisition, making duration, predicted future activity, and conversational context important for distinguishing short participation from sustained transfer. Under continuous discovery, NTT recall is 0.607 compared with 0.470 for ITT, consistent with the greater ambiguity of detecting floor transfer during competitive overlap.

Timing analysis further shows that NTT recall is highest at moderate positive gaps and declines as the gap approaches 1~s, while ITT recall decreases with deeper overlap. Longer gaps can resemble within-speaker pauses or unresolved floor opportunities, whereas deeper overlap increases ambiguity between floor transfer and competing listener activity. Approximately 85\% of strict continuous false positives arise from background/other candidates rather than near known BC or BI events. Therefore, BC and BI are prominent confusions at supplied decision points. However, most continuous false alarms arise from the broader search space of peaks, pauses, and automatically derived boundaries that do not correspond to labelled interaction events.

\subsection{Augmentation Effects}

Table~\ref{tab:aug} shows that augmentation primarily shifts
the precision--recall operating point. Pre-generated
background-bed mixing yields the highest precision but lowest
recall, whereas synthetic augmentation increases recall
relative to no augmentation. The full condition gives the
highest nominal F1; however, all paired $\Delta$F1 confidence
intervals relative to no augmentation include zero.
These differences indicate changes in operating behaviour
rather than a statistically significant improvement in F1.

\begin{table*}[t]
\caption{Effect of training-time augmentation on VoxConverse
Task~C at $\pm 0.50$~s. Paired $\Delta$F1 is measured against
no augmentation over 214 common TEST sessions.}
\label{tab:aug}
\centering
\scriptsize
\begin{tabular}{lrrrrl}
\toprule
Condition & P & R & F1 & $\Delta$F1 & 95\% CI for $\Delta$F1 \\
\midrule
No augmentation & 0.567 & 0.516 & 0.541 & -- & -- \\
Synthetic only & 0.543 & 0.556 & 0.549 & +0.0077 & [-0.0050, 0.0203] \\
Pre-generated background beds only & \textbf{0.617} & 0.484 & 0.542 & +0.0015 & [-0.0086, 0.0128] \\
Full & 0.536 & \textbf{0.569} & \textbf{0.552} & +0.0109 & [-0.0016, 0.0232] \\
\bottomrule
\end{tabular}
\end{table*}

\subsection{Incoming-Speaker Identification}

Table~\ref{tab:speakerdiag} evaluates post-boundary incoming-speaker
identification under progressively less reference-dependent speaker
representations. A speaker prototype is an aggregate embedding constructed
from available speech: the reference condition uses RTTM speaker labels,
the automatic full-recording condition uses diarised speech from the entire
recording, and the prefix-enrolment condition restricts prototype construction
to diarised speech preceding the decision time. Conditional identification
accuracy is computed only over temporally matched transitions for which a
speaker identity is available, whereas identification coverage is the
proportion of matched transitions for which an identity can be produced.
Event+speaker F1 is stricter: a prediction is credited only when the
transition is matched within the primary $\pm0.50$~s tolerance and the
incoming speaker is also identified correctly.

Performance decreases when reference-derived speaker
representations are replaced by automatic diarization and
decreases further when enrolment is restricted to preceding
speech (Table~\ref{tab:speakerdiag}). This reflects two
sources of difficulty: automatic clustering introduces speaker
identity uncertainty, while past-only enrolment provides less
evidence for speakers with limited prior speech. Since all
three conditions use post-boundary query audio, the results
evaluate incoming-speaker identification rather than
pre-transition next-speaker prediction.

\begin{table}[t]
\caption{Post-boundary incoming-speaker identification on VoxConverse TEST at
the primary $\pm0.50$~s tolerance. Event+speaker F1 requires both a correctly
matched transition and the correct incoming-speaker identity.}
\label{tab:speakerdiag}
\centering
\scriptsize
\setlength{\tabcolsep}{2.5pt}
\renewcommand{\arraystretch}{1.12}
\begin{tabularx}{\columnwidth}{@{}Xccc@{}}
\toprule
Speaker representation &
\shortstack{Cond. ID\\acc.} &
\shortstack{ID\\coverage} &
\shortstack{Event+speaker\\F1} \\
\midrule
RTTM reference prototypes
& 0.871 & 1.000 & 0.481 \\

Automatic full-recording prototypes
& 0.810 & 0.988 & 0.442 \\

Automatic prefix-enrollment prototypes
& 0.721 & 0.955 & 0.380 \\
\bottomrule
\end{tabularx}
\end{table}

\subsection{Speaker Count and Overlap}

Fourteen TEST recordings with fewer than two RTTM speakers
are excluded only from the multiparty correlation analyses;
they remain part of the main Task~C evaluation. Session F1
does not vary monotonically across recordings with two,
three, four, five, or six or more speakers, and speaker count
is not significantly associated with session F1
($\rho=0.069$, $p=0.310$). Speaker count alone therefore
does not appear to explain the observed variation in
continuous-discovery performance.

Overlap shows a different pattern. Although F1 across the
low-, medium-, and high-overlap tertiles is non-monotonic
(0.587, 0.631, and 0.503), the continuous overlap ratio
has a modest negative association with session F1 across
217 multiparty recordings ($\rho=-0.159$, $p=0.019$).
This is consistent with the lower recall observed for
interruption transfers and deeper overlap, while indicating
that overlap is associated with, rather than solely
determining, interaction difficulty.

\section{Multi-TPC Evaluation and Robustness Analysis}
\label{sec:multitpc}
Multi-TPC is evaluated under three complementary protocols. First, we report the original group-disjoint held-out evaluation using the conversational group reserved before model selection. Second, we assess cross-group robustness through a post-hoc nested leave-one-group-out evaluation across all seven conversational groups, with model and operating-point selection confined to the corresponding inner folds. Third, we use a fixed-configuration leave-one-group-out study to examine architectural and modality ablations. The original held-out evaluation is retained as the confirmatory result, whereas the seven-group analyses provide post-hoc robustness and explanatory evidence rather than a second untouched test.

\subsection{Event Definition and Evaluation Protocol} Multi-TPC events are derived from the native 60-Hz participant-activity timeline. A normal turn transfer (NTT) is anchored at outgoing-speaker floor release when a different participant subsequently acquires the floor after a gap of at least 0.20~s. An interruption turn transfer (ITT) is anchored at incoming-speaker onset when the incoming participant begins before the incumbent releases, the incumbent resolves within 1.0~s, and the incoming participant sustains primary speech for at least 0.30~s after release. Alternative incumbent interpretations of the same physical transfer are collapsed into a single event. Future-activity annotations may provide training targets, but future ground-truth activity is not available at inference. 

Two evaluation protocols are kept distinct. The original group-disjoint evaluation uses five conversational groups for training, one for validation, and one pre-reserved group for held-out evaluation. The held-out group is not used for model or operating-point selection. After this evaluation had been completed, cross-group robustness was examined separately using post-hoc nested leave-one-group-out evaluation across all seven groups. In each outer fold, the evaluation group remains excluded while model and operating-point selection use only group-wise inner folds from the remaining data; automatic comparators are refitted under the same protocol. 

The primary event-matching tolerance is $\pm0.50$~s, with $\pm0.70$ and $\pm1.00$~s reported as localisation-sensitivity analyses. For the seven-group analysis, the inner procedure selects the operating point using $\pm1.00$~s, while the held-out outer group is never used for selection. Consequently, the original group-disjoint experiment provides the held-out evaluation, whereas the later seven-group analysis provides post-hoc cross-group robustness evidence rather than a second untouched test.

\subsection{Original Group-Disjoint Held-Out Evaluation}

Table~\ref{tab:mtpcr11} reports the original held-out evaluation under the
common Multi-TPC protocol. Among the fully automatic methods, the proposed
dual-process system has the highest observed Task~C F1 at all three matching
tolerances, whereas Triadic VAP attains the highest recall at
$\pm1.00$~s. The annotation-assisted Lee \emph{et al.} condition achieves
higher Task~C F1 under its stronger current-state information contract and is
therefore shown separately rather than treated as an equal-input automatic
comparison. For the proposed system, the median absolute localisation error among matched events at $\pm1.00$~s is 0.407~s. Because the evaluation uses a single
held-out conversational group, these differences describe performance on the
reserved group and do not establish cross-group method ordering.

\begin{table*}[t]
\caption{Original group-disjoint Multi-TPC held-out evaluation.
Tasks~A/B report Macro-F1; Task~C reports event F1 at three matching
tolerances and precision/recall at $\pm1.00$~s. The annotation-assisted
Lee \emph{et al.} condition uses a stronger input contract.}
\label{tab:mtpcr11}
\centering
\scriptsize
\setlength{\tabcolsep}{3.5pt}
\renewcommand{\arraystretch}{1.10}
\begin{tabular}{lccccccc}
\toprule
Method &
A Macro-F1 &
B Macro-F1 &
C F1 $\pm.50$ &
C F1 $\pm.70$ &
C P $\pm1.00$ &
C R $\pm1.00$ &
C F1 $\pm1.00$ \\
\midrule
Proposed dual-process
& 0.500 & 0.563 & 0.260 & 0.352 & 0.363 & 0.576 & 0.445 \\

Lee \emph{et al.}~\cite{lee2024eot} (automatic)
& 0.324 & 0.462 & 0.250 & 0.307 & 0.286 & 0.619 & 0.391 \\

Triadic VAP~\cite{triadicvap}
& 0.307 & 0.395 & 0.215 & 0.313 & 0.309 & 0.706 & 0.430 \\

MuVAP~\cite{muvap}
& 0.457 & 0.396 & 0.238 & 0.315 & 0.276 & 0.562 & 0.371 \\
\midrule
Lee \emph{et al.}~\cite{lee2024eot} (annotation-assisted)
& 0.399 & 0.598 & 0.345 & 0.463 & 0.685 & 0.460 & 0.551 \\
\bottomrule
\end{tabular}
\end{table*}

\subsection{Post-hoc Seven-Group Cross-Group Evaluation}

Table~\ref{tab:mtpcr12} reports the post-hoc nested evaluation
across all seven conversational groups. The literature-based
comparators are Multi-TPC adaptations refitted under the same
outer-fold protocol. The observed ordering differs from the
original held-out evaluation: Triadic VAP has the highest mean
Task~C F1 at all three tolerances, while MuVAP and the proposed
system remain close at the broader tolerances. The proposed
system nevertheless has higher mean Task~C F1 than the
automatic Lee \emph{et al.} comparator across all tolerances.

These differences are descriptive. No paired outer-fold
comparison remains significant after Holm correction. For the
proposed system, the group-clustered 95\% bootstrap interval
for Task~C F1 is [0.317, 0.371] at $\pm0.50$~s and
[0.469, 0.538] at $\pm1.00$~s. At $\pm1.00$~s, the Triadic-VAP
comparison has an unadjusted paired Wilcoxon $p=0.0469$,
but the Holm-adjusted value is $p=0.4688$. The change in
method ordering relative to the single held-out group
highlights the importance of cross-group evaluation without
establishing a statistically supported ranking among the
automatic methods.

\begin{table*}[t]
\caption{Post-hoc seven-group nested leave-one-group-out
evaluation (mean $\pm$ SD across outer groups).
Tasks~A and B report Macro-F1, with Task~B using
$\pm 1.00$~s candidate labels. Task~C reports event F1
across matching tolerances and precision/recall at
$\pm 1.00$~s.}
\label{tab:mtpcr12}
\centering
\scriptsize
\setlength{\tabcolsep}{3.1pt}
\begin{tabular}{lccccccc}
\toprule
Method &
A Macro-F1 &
B Macro-F1 &
C F1 $\pm.50$ &
C F1 $\pm.70$ &
C P $\pm1.00$ &
C R $\pm1.00$ &
C F1 $\pm1.00$ \\
\midrule
Proposed dual-process
& 0.572$\pm$0.045 & 0.607$\pm$0.021
& 0.343$\pm$0.040 & 0.439$\pm$0.045
& 0.387 & 0.747 & 0.504$\pm$0.050 \\

Lee \emph{et al.}~\cite{lee2024eot} (automatic)
& 0.549$\pm$0.008 & 0.532$\pm$0.014
& 0.257$\pm$0.029 & 0.347$\pm$0.037
& 0.317 & 0.693 & 0.428$\pm$0.044 \\

Triadic VAP~\cite{triadicvap}
& 0.594$\pm$0.068 & 0.599$\pm$0.043
& 0.382$\pm$0.065 & 0.474$\pm$0.076
& 0.467 & 0.701 & 0.553$\pm$0.076 \\

MuVAP~\cite{muvap}
& 0.658$\pm$0.029 & 0.622$\pm$0.033
& 0.346$\pm$0.032 & 0.448$\pm$0.035
& 0.404 & 0.744 & 0.516$\pm$0.041 \\
\bottomrule
\end{tabular}
\end{table*}

\subsection{Fixed-Configuration Ablation Analysis}
\begin{figure*}[t]
\centering
\includegraphics[width=0.92\textwidth]{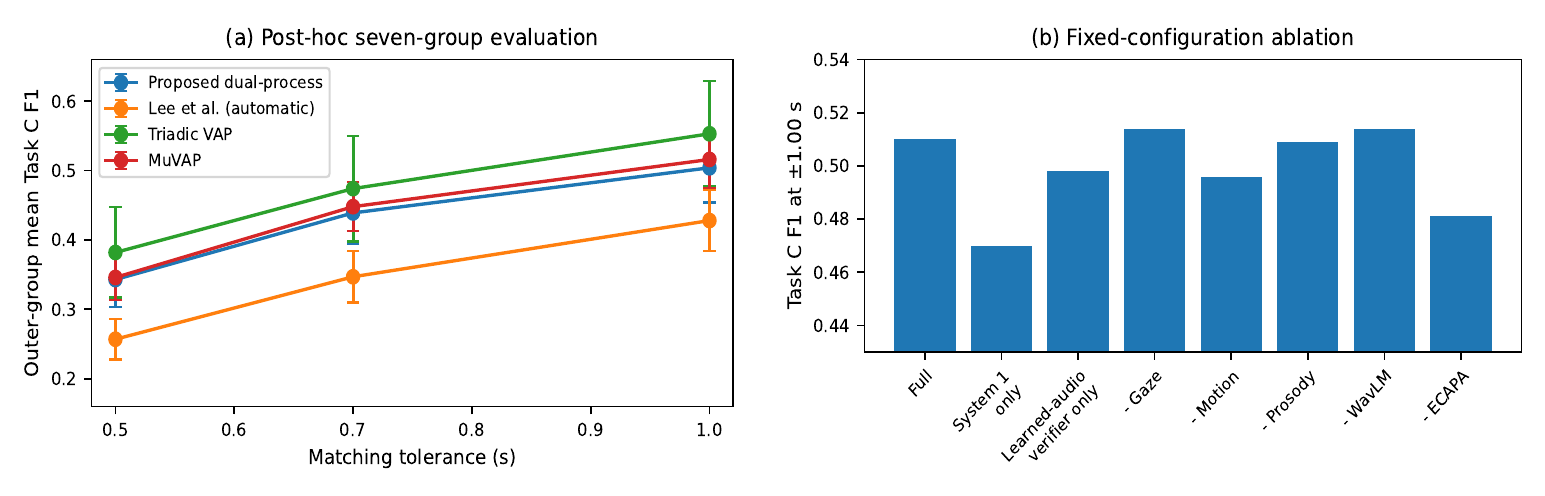}
\caption{Multi-TPC cross-group and ablation results. (a) Post-hoc
seven-group Task~C F1 across matching tolerances for the four fully automatic
methods; error bars denote outer-group standard deviation. (b)
Fixed-configuration Task~C ablation at $\pm1.00$~s; the Full configuration
is the paired reference for the ablation analysis.}
\label{fig:multitpcresults}
\end{figure*}

Table~\ref{tab:mtpcablation} compares the full system with stage-level
restrictions and individual evidence-family removals under the same
seven-group evaluation. The full configuration has higher mean Task~C F1
than both the System~1-only and learned-audio-only configurations at all
three tolerances, consistent with complementary contributions from the fast
candidate representation and deliberative evidence. Figure~\ref{fig:multitpcresults}(b)
summarizes the corresponding $\pm1.00$~s Task~C ablation pattern.

The individual ablations are less uniform. Removing ECAPA produces the
largest reduction at $\pm1.00$~s, with a smaller reduction from removing
motion, whereas gaze, prosody, and WavLM have small or tolerance-dependent
effects. No ablation comparison remains significant after Holm correction
over the reported test family. The results therefore support the
architecture-level combination descriptively, but do not establish that
every individual modality contributes an independent improvement.

\begin{table*}[t]
\caption{Fixed-configuration Multi-TPC ablation across seven outer groups
(mean $\pm$ SD). The Full configuration is the paired reference for this
ablation analysis; $\Delta_{1.0}$ denotes Full minus the ablated configuration
for Task~C at $\pm1.00$~s.}
\label{tab:mtpcablation}
\centering
\scriptsize
\setlength{\tabcolsep}{3.0pt}
\begin{tabular}{lcccccc}
\toprule
Configuration &
A Macro-F1 &
B Macro-F1 &
C F1 $\pm.50$ &
C F1 $\pm.70$ &
C F1 $\pm1.00$ &
$\Delta_{1.0}$ \\
\midrule
Full
& 0.580$\pm$0.039
& 0.604$\pm$0.024
& 0.342$\pm$0.037
& 0.442$\pm$0.043
& 0.510$\pm$0.048
& -- \\

System~1 only
& 0.580$\pm$0.039
& 0.572$\pm$0.033
& 0.316$\pm$0.043
& 0.409$\pm$0.039
& 0.470$\pm$0.048
& +0.040 \\

Learned-audio verifier only
& 0.580$\pm$0.039
& 0.605$\pm$0.019
& 0.328$\pm$0.042
& 0.429$\pm$0.043
& 0.498$\pm$0.043
& +0.012 \\

$-$ Gaze
& 0.578$\pm$0.047
& 0.607$\pm$0.020
& 0.342$\pm$0.044
& 0.441$\pm$0.049
& 0.514$\pm$0.050
& $-$0.004 \\

$-$ Motion
& 0.564$\pm$0.066
& 0.604$\pm$0.018
& 0.339$\pm$0.040
& 0.430$\pm$0.044
& 0.496$\pm$0.054
& +0.014 \\

$-$ Prosody
& 0.565$\pm$0.044
& 0.610$\pm$0.017
& 0.346$\pm$0.043
& 0.439$\pm$0.046
& 0.509$\pm$0.046
& +0.001 \\

$-$ WavLM
& 0.580$\pm$0.039
& 0.598$\pm$0.028
& 0.334$\pm$0.037
& 0.440$\pm$0.046
& 0.514$\pm$0.052
& $-$0.004 \\

$-$ ECAPA
& 0.580$\pm$0.039
& 0.584$\pm$0.022
& 0.337$\pm$0.037
& 0.430$\pm$0.046
& 0.481$\pm$0.052
& +0.029 \\
\bottomrule
\end{tabular}
\end{table*}

The Full row is the paired reference for this ablation analysis and is not
used to replace the headline seven-group result in
Table~\ref{tab:mtpcr12}.

\section{Discussion}

\subsection{Discovery, Verification, and Action Interpretation}

The main empirical separation is between interpreting a supplied decision
point and discovering a floor transfer in continuous interaction.
Although System~1 places a candidate near most VoxConverse reference
transfers, Task~C remains substantially harder than supplied-point
classification (Table~\ref{tab:mainresults}). The dominant difficulty is
therefore not candidate availability alone, but selecting, rejecting, and
localising plausible transition hypotheses in continuous speech.

This is the central methodological finding: supplying a boundary hides the search and false-opportunity burden that a deployed event detector must bear. The same-data silence baselines reinforce this interpretation (Table~\ref{tab:baselines}). Silence alone produces many false alarms,
whereas requiring speaker change improves selectivity but remains poorly
suited to interruptions that occur without mutual silence. The event-type
analysis further separates semantic and temporal errors: brief listener
actions are difficult HOLD cases at supplied decision points, while most
continuous false positives arise from the broader background of automatically
generated non-event candidates. Continuous turn-taking therefore requires
both action interpretation and temporal search.

\subsection{Overlap, Future Activity, and Evidence Fusion}

Speaker count is not significantly associated with VoxConverse
session F1, whereas overlap is negatively associated with
performance and interruptions are harder to detect than normal
transfers. A post-hoc analysis of frozen \taskC{} TEST outcomes
found lower event F1 in non-English-dominant recordings (0.488)
than in English-dominant recordings (0.645) under conservative
cross-model language agreement. The difference remained after
accounting descriptively for recording duration, overlap, and
speaker count. However, incomplete language-label coverage
and differences in programme and conversational structure
prevent attributing the performance gap to language alone.

Future-activity prediction provides anticipatory evidence but
does not itself establish a floor transfer: another speaker
may become active without acquiring the floor, while the
current speaker may pause without yielding. System~2
therefore combines future-activity cues with acoustic,
linguistic, speaker-aware, and temporal evidence.
Component comparisons and Multi-TPC ablations further show
that individual evidence sources can be complementary,
redundant, or condition-dependent.

\subsection{Cross-Group Robustness and Claim Boundaries}

The Multi-TPC results show that method ordering depends on the conversational
group and evaluation protocol. In the original group-disjoint held-out
evaluation, the proposed system has the highest observed Task~C F1 among the
fully automatic common-protocol methods. In the subsequent post-hoc
seven-group analysis, however, Triadic VAP has the highest mean Task~C F1,
with MuVAP also close to the proposed system
(Table~\ref{tab:mtpcr12} and Fig.~\ref{fig:multitpcresults}(a)).
No paired comparison remains statistically significant after Holm correction, so these differences should be interpreted as descriptive cross-group variation rather than a statistically supported method ranking.

The fixed-configuration ablation provides a complementary architectural
result (Table~\ref{tab:mtpcablation} and
Fig.~\ref{fig:multitpcresults}(b)). The full configuration consistently
outperforms the System~1-only and learned-audio-only restrictions in Task~C,
while individual evidence removals have smaller and less uniform effects.
ECAPA shows the clearest positive contribution at the broad tolerance, with
motion providing a smaller effect; gaze, prosody, and WavLM are weak or
tolerance-dependent in this configuration. Because no ablation comparison remains statistically significant after Holm correction, these modality effects remain descriptive rather than evidence that every source is independently necessary.

The incoming-speaker analysis has a separate information boundary.
Identification quality decreases as reference-derived speaker information is
removed, but all reported conditions still use post-boundary query speech.
They therefore evaluate incoming-speaker identification after a detected
transition, not causal prediction of who will speak next.

Finally, the dual-process interpretation is functional rather than
psychological. System~1 is optimised for high-recall opportunity discovery,
whereas System~2 applies richer reasoning only to the resulting candidate
set. The measured reduction in verifier invocations supports this selective
computation design, but does not imply cognitive equivalence or end-to-end
real-time operation.

\section{Reproducibility and Limitations}

VoxConverse model, threshold, post-processing, and timing
selection use TRAIN/VAL only; TEST is reserved for final
evaluation. Results use the canonical 6,348-event TEST set.
Uncertainty is estimated using session-level bootstrap resampling,
with 1,000 replicates for the primary Task~C confidence interval
and 2,000 for paired augmentation comparisons. Frozen
configurations and evaluation scripts are retained in the
project repository.

A key VoxConverse limitation is that turn-taking events are
derived from speaker-diarization annotations rather than
native conversational-action labels. The 0.50-s brief-response
threshold is task-specific, and the 300-clip manual audit
assesses action labels rather than event timing. Waveform
augmentation and background-bed mixing follow documented
training procedures, although their perceptual effects on
event-label validity were not independently assessed.
The executed VoxConverse pipeline uses bounded look-ahead
and offline-precomputed diarization, ASR, embeddings, and
caches. Runtime measurements are therefore component-level rather than
end-to-end. Incoming-speaker identification also relies on
post-boundary speech.

Multi-TPC contains seven conversational groups and 23
complete locally audited sessions; one motion stream is
excluded due to a severe duration mismatch, and prosody
is incomplete in two sessions. The seven-group nested
evaluation is post-hoc, and no fixed-configuration ablation
comparison survives Holm correction. Modality effects
therefore remain descriptive and potentially group-dependent.
Synchronised gaze, motion, and prosody are available only
in Multi-TPC.

\section{Conclusion}

This work develops and evaluates an event-triggered, dual-process-inspired
formulation of multiparty floor-transfer detection. Its contribution is
not a new acoustic encoder or a claim that continuous prediction is new,
but an explicit separation of finding candidate moments, deciding whether
those moments represent a floor transfer, and localising accepted events. Across VoxConverse, high candidate coverage shows that the main
difficulty lies in rejecting and localising plausible transition hypotheses,
particularly under overlap and interruption, rather than simply generating
more candidates.

Multi-TPC further shows that method ordering varies across conversational
groups, while fixed-configuration ablations support combining fast candidate
discovery with richer deliberative evidence. Overall, the results support a
dual-process-inspired decomposition in which high-recall opportunity discovery
is followed by selective interpretation using complementary acoustic,
linguistic, speaker-aware, future-activity, and multimodal cues. Remaining
work includes broader human validation of derived events, integrated
streaming evaluation, and genuinely pre-transition next-speaker prediction.

\bibliographystyle{IEEEtran}
\bibliography{citations}

\end{document}